\documentclass[aps, prl, showpacs, floatfix,twocolumn]{revtex4}

\usepackage{graphicx}
\usepackage{graphicx}
\usepackage {amssymb}
\usepackage {amsmath}
\usepackage[bbgreekl]{mathbbol}

\newcommand{\vect}[1]{{\mathrm {\mathbf #1}}} 

\DeclareMathOperator{\diag}{diag}

\newcommand{\reffig}[1]{Fig.~\ref{#1}}

\begin{document}

\title{Scar-like structures and their localization in a perfectly
  square optical billiard}

\author{I. Babushkin}
\affiliation{Weierstrass Institute for Applied Analysis and Stochastics
Mohrenstr. 39, 10117,  Berlin, Germany}
\email{babushkin@wias-berlin.de}

\begin{abstract}
We show that scar-like structures (SLS) in a wide aperture vertical
cavity surface emitting laser (VCSEL) can be formed even in a
perfectly square geometry due to interaction of polarization and
spatial degrees of freedom of light. We show also that dissipation in
the system induces an order among the cavity modes, so that SLS become
preferred at lasing threshold. More generally, modes which are more
localized both in coordinate and momentum space have in average lower
losses.
\end{abstract}



\maketitle


Quantum billiards traditionally attract a strong attention
\cite{stoeckmann99, guhr98} in the quantum chaos studies. Different
types of systems, ranging from acoustic and microwave resonators to
optical cavities and quantum dots belong to that class.  In most
quantum billiards waves of certain (not necessary quantum) nature
freely move in the region of certain shape, surrounded by reflective
boundaries.  Such systems are described by an energy operator $\hat H$
proportional to Laplacian $\hat H \sim \Delta \equiv \partial_{xx} +
\partial_{yy}$ (in 2D case), supplied with corresponding boundary
conditions.

In contrast to billiards, in quantum systems with many internal
degrees of freedom quantum chaos arises even for trivial boundary
conditions (if one can speak about boundary conditions at all) due to
complex structure of $\hat H$.  To such class belong nuclei as well as
other many-body systems \cite{guhr98,ullmo07}.  We will refer to the
later systems as to ``operator-determined'' whereas simple billiards
will be called ``boundary-determined'' ones.

Some systems however belong to an intermediate type, for example
billiards having a ``ball'' with nontrivial internal structure
(i.~e. possessing internal degrees of freedom) interacting, in one or
another way, with its kinetic motion.  Up to now only few such systems
are known, among them are quantum dots in presence of spin-orbital
coupling \cite{berggren01,novaes04}, as well as anisotropic acoustic
cavities \cite{schaadt99}.  In contrast, ``photon billiards'' (i.~e.
optical and microwave cavities) are traditionally considered as fully
``boundary-determined'' ones \cite{stoeckmann99}.

One important type of an optical billiard is a vertical cavity surface
emitting laser (VCSEL). Recent advances in technology allowed to
produce wide aperture, highly homogeneous devices of arbitrary shape
\cite{huang02,gensty05,chen03c,chen09}.  In square devices, one of the
most prominent features is the presence of scar-like structures (SLS)
\cite{huang02,chen03d,babushkin08b,schulz-ruhtenberg08,chen02a,chen03a},
which are localized along classical trajectories.  If we consider
VCSEL as a ``boundary-determined'' billiard, appearance of such
structures must be attributed to some other mechanism such as
deformations of the boundaries \cite{huang02} or mode-locking
\cite{chen02a}. This was done however without a rigorous verification.

In the present article we show that, in fact, the presence of SLS does
not require any disturbance of the square boundaries.  Some amount of
nonintegrability is provided by a coupling of internal (light
polarization) and transverse degrees of freedom of photons, which we
call here polarization-transverse coupling (PTC), appearing due to
presence of direction-depended anisotropy created by the cavity
mirrors \cite{babushkin08b}.

Because fully integrable billiard is separable, the presence of SLS
points out to a deviation from the complete integrability.  This
deviation can be regulated by misalignment of the intracavity
anisotropy to the boundaries. However, it is always relatively small,
making the situation similar to quasintegrable scalar billiards
\cite{bogomolny04,bogomolny06}.

Moreover, we show that the complicated modes created by PTC are
ordered in presence of dissipation, so that SLS become 
preferable (by having less losses) at lasing threshold.  More
generally, we demonstrate that the modes close to threshold are more
localized, in average, both coordinate and momentum space, comparing
to the modes with higher losses.  This ordering allows to explain why
SLS so naturally appear in VCSELs.

Similar relation between localization and dissipation was found very
recently for a fully chaotic system \cite{ermann09}. Localization of
long-lived modes appearing in the vicinity of avoided level crossing
(ALC) was also pointed out (for the coordinate space only) for
dielectric microcavities \cite{wiersig06}.  However, in contrast to
\cite{wiersig06}, in our case the existence of SLS is neither directly
related to ALC, nor to dissipation (i.~e. to connection to the outer
world).

Using our system as an example we also demonstrate that the measures of
localization in coordinate and momentum space are very sensitive to
even a small deviation from the complete integrability.


Despite of sufficiently nonlinear nature of the lasing process, many
properties of the spatio-temporal distribution in broad-area VCSELs
can be grasped already in a linear approximation \cite{babushkin08b}.
Although the working area of VCSEL can be maid very homogeneous in the
transverse direction, the longitudinal structure of the cavity is
rather complicated.  In particular it includes the multilayered
structure playing the role of the cavity mirrors (so called
distributed Bragg mirrors, DBRs).  However, the longitudinal degree of
freedom can be excluded from the description \cite{loiko01} in an
effective way due to single longitudinal mode operation of the device.
As a result of such reduction and of subsequent linearization near the
lasing threshold \cite{babushkin08b} the cavity structure is described
by a single linear operator defining evolution of the complex vector
field envelope $\vect E(\vect r_\bot,t)$ with time:
\begin{equation}
  \label{eq:eh}
  \dot {\vect E}(\vect r_\bot,t) = i \hat H \vect E (\vect r_\bot,t),
\end{equation}
where the dot means the partial time derivative, $\vect r_\bot =
\{x,y\}$ are the transverse coordinates and $\hat H$ is a linear
operator acting on a transverse field distributions $\vect E(\vect
r_\bot)$.  $\hat H$ is most easily described for a transversely
infinite VCSEL.  In this case, its eigenfunctions are tilted waves
$\vect E \sim e^{-i\vect r_\bot \vect k_\bot}$ with certain transverse
wavevector $\vect k_\bot = \{k_x,k_y\}$. Therefore, $\hat H$ can be
written in the transverse Fourier space $\vect k_\bot =\{k_x,k_y\}$ as
a multiplication to a $2 \times 2$ matrix-function $\beta_\infty(\vect
k_\bot)$: $\beta_\infty= a k_\bot^2 + \Gamma + bs(k_\bot) + i\kappa
\Upsilon(k_\bot)$.  Here $a$, $b$ and $\kappa$ are some constants
defined by the parameters of the device, $k_\bot = |\vect k_\bot|^2$
describes a free kinetic motion of the light inside the cavity,
$\Gamma$ is the intracavity anisotropy, which in the Cartesian basis
formed by principal anisotropy axis can be written as $\Gamma =
\diag(\gamma_p+i\gamma_a, -\gamma_p-i\gamma_a)$, where
$\diag(\cdot,\cdot)$ is a $2\times 2$ diagonal matrix with
corresponding elements on the diagonal, $\gamma_p$ and $\gamma_a$ is
the phase and amplitude anisotropies.  The matrices $s(\vect k_\bot)$
and $\Upsilon(\vect k_\bot)$ represent the $\vect k_\bot$-dependent
phase and amplitude anisotropy, created by DBRs \cite{loiko01,
  babushkin08b}.  The main axes of this anisotropy are perpendicular
and parallel to $\vect k_\bot$.  Importantly, $\Upsilon$ contains the
outcoupling losses as well as the gain.  For the mode at threshold the
losses and gain exactly compensate each other.


In the transverse directions the light in VCSEL is guided by a thin
oxide aperture.  Under certain approximation (in particular assuming
the ideal reflection at the side boundaries), the modes of such
waveguide are the functions of the type $\vect E \cos(k_xx)
\cos(k_yy)$ which contain four spots with equal amplitudes and
polarization directions in $\vect k_\bot$-space.  In contrast,
eigenmodes of DBRs have polarization either perpendicular or parallel
to $\vect k_\bot$, which can not be represented by any combination of
waveguide modes with a fixed $k_\bot$.  Therefore DBR reflection
unavoidly rescatters the eigenmodes of the waveguide into the ones
with different $ k_\bot$, which creates PTC \cite{babushkin08b}.

Using $\beta_\infty$ and the above mentioned properties of the
waveguide modes, one can directly construct an operator $\beta_s$,
which represents $\hat H$ in basis of waveguide modes (see
\cite{babushkin08b} for details), as well as its reduction  $\beta_p$
with  completely neglected PTC:
\begin{equation}
  \label{eq:s-act}
  E^{(i)}_{km} = \sum_{j,l,n}\beta_s^{ijklmn} E^{(j)}_{ln}; \; 
  \beta_p^{ijklmn} = \delta_{kl}\delta_{nm} \beta_s^{ijklmn}.
\end{equation}
Here $E^{(j)}_{nm}$ is the $j$th polarization of the transverse mode
$n,m$ of the waveguide and $\delta_{nm}$ is the Kronecker
$\delta$-symbol.


\begin{figure}[tpb]
  \includegraphics[width=0.45\textwidth,clip=]{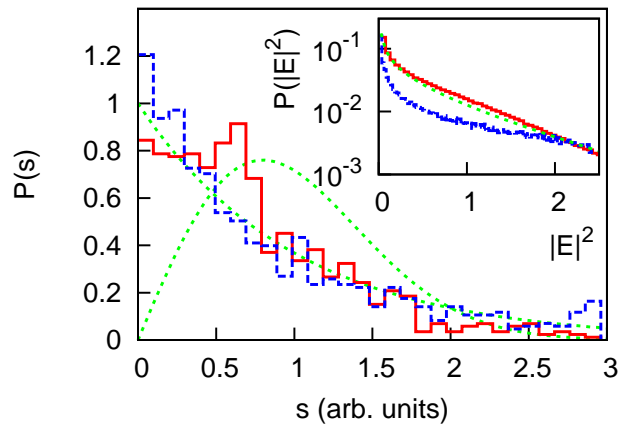} 
  \caption{ \label{fig:stat} (Color online) Statistics of the
    eigenvalues of $\beta_s$ (red solid curve) and $\beta_p$ (blue
    dot-dashed curve) for $\alpha = \pi/15$.  Dotted green lines shows
    Poissonian and Wigner distributions. In the inset, the statistics
    $P(|\vect E|^2)$ is shown (with the same denotations as in the
    main figure; dotted green line shows Porter-Thomas distribution).}
 \end{figure}


 For numerical computation of the eigenvalues and eigenfunctions of
 $\beta_s^{ijklmn}$ it is transformed into a square matrix by
 introducing indices $I=i+2 k+2
 (N_{\mathrm{max}}-N_{\mathrm{min}}+1)m$, $J=j+2 l+2
 (N_{\mathrm{max}}-N_{\mathrm{min}}+1) n$, where $N_{\mathrm{max}}$
 and $N_{\mathrm{min}}$ are the maximal and minimal waveguide mode
 numbers, defining a cut-off for high- and low- order modes.
 Physically, cut-off for high-order modes is necessary because they
 are not guided anymore in transverse direction.  On the other hand,
 if we consider the structures formed from the modes which are
 sufficiently far from $k_\bot=0$, very low order modes can be also
 neglected.  For the simulations the values of $N_{\mathrm{max}}=30$,
 $N_{\mathrm{min}}=10$ were taken.  In this case the matrix
 $\beta_s^{I,J}$ has the size $\sim 1000\times 1000$.  Typical VCSEL
 parameters were used in simulations, in particular, the intracavity
 anisotropy $\gamma_a = 0.1$ ns$^{-1}$, $\gamma_p = 30$ ns$^{-1}$ were
 taken.  The detuning $\delta$ of the cavity resonance from the gain
 line center (which enters to $\Upsilon$) controls the values of
 $k_\bot$ which have maximal gain \cite{sanmiguel95b}. It is chosen
 large enough so that the modes with $k_\bot$ above the high-$k_\bot$
 cut-off are preferred in the infinite device.  Due to presence of the
 cut-off, $\vect k_\bot$ near diagonal $|k_x|=|k_y|$ are selected at
 threshold \cite{babushkin08b, schulz-ruhtenberg08}.


 The statistics $P(s)$ of nearest neighbor separation $s_i\sim
 E_{i+1}-E_i$ of the eigenvalues $E_i$ of the matrices $\beta_s$ and
 $\beta_p$ is presented in \reffig{fig:stat} in comparison to
 Poissonian ($P(s)=e^{-s}$) and Wigner ($P(s) = \frac{\pi}{2}e^{-\pi
   s^2/4}$) distributions.  The anisotropy axes in \reffig{fig:stat}
 are assumed to be rotated to a small angle $\alpha= \pi/15$ in
 respect to $x$-axis (such small displacement may also exist in real
 devices \cite{schulz-ruhtenberg08}, although in tendency the
 anisotropy is aligned to the boundaries). In this case, noticeable
 deviation from the Poissonian statistic is present for $\beta_s$.
On the other hand, the eigenvalues of $\beta_p$ obey Poissonian
statistics for every $\alpha$.  In contrast, for $\alpha =0$, the
deviation of $P(S)$ from the Poissonian distribution for $\beta_s$ is
not noticeable anymore ( not shown in \reffig{fig:stat}).
This shows, that although PTC itself plays a critical role in the
statistics of eigenvalues, the alignment of the intracavity
anisotropy  to the boundaries is also important, as it increases the
degree of mode mixing produced by PTC.

\begin{figure}[tpb]
\includegraphics[width=0.45\textwidth,clip=]{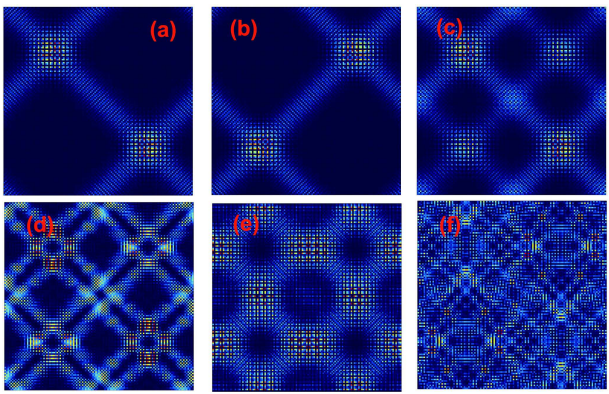}
\caption{ \label{fig:scheme} (Color online) Example of scared (a)--(e)
  as well as non-localized (f) eigenfunctions of $\beta_s$. In
  (a)--(c) $|E_x|^2$, $|E_y|^2$ and the intensity $I$ of the
  eigenfunction at threshold (i.~e. having the lowest losses) are
  shown; in (d)--(f) the intensity $I$ distribution of some subsequent
  modes is presented. }
\end{figure}

Another important property of the operator $\beta_s$ (both for $\alpha
> 0 $ and $\alpha =0$) is the presence of eigenfunctions localized
along classical trajectories (i.~e. SLS).  Examples of such structures
are shown in \reffig{fig:scheme}(a)-(e).  In particular, in
\reffig{fig:scheme}(a)-(c) the amplitudes of $x$- and $y$-polarization
components as well as the full intensity of the mode at threshold
(i.~e. one with lowest losses) are shown.  The mode in
\reffig{fig:scheme}(c) is localized along tree different classical
trajectories. It is interesting that particular polarization
components, in contrast to the full intensity, are not localized along
the complete trajectory in this case.  Similar phenomena were observed
also experimentally \cite{schulz-ruhtenberg08} (cf. also
\cite{chen03a} where different polarizations follow different
trajectories).

In general, SLS are very common for $\beta_s$, for both $\alpha = 0$
and $\alpha > 0$ (see some further examples in
\reffig{fig:scheme}(d),~(e)).  As a rule, they contain more than one
classical trajectory.  Like \cite{bogomolny04,bogomolny06}, SLS in our
system seemingly do not become more rear with increasing of $k_\bot$.
On the other hand, far from threshold the eigenfunctions becomes less
localized (see an example in \reffig{fig:scheme}(f)).

The existence of SLS clearly points out to deviation from the full
integrability, because the later assumes an existence of coordinate
system where the billiard becomes fully separable, which excludes the
possibility of SLS. For $\alpha =0$ this deviation is ``undetected'' by
$P(s)$.

In contrast to $\beta_s$, the set of eigenfunctions of $\beta_p$ do
not contain SLS. For $\alpha =0$, $\beta_p$ is diagonal.  For $\alpha
\ne 0$ it consists of $2\times 2$ blocks at the main diagonal
(describing the polarization degrees of freedom). Therefore, all the
cases we consider can be arranged in order of increasing of
``nonintegrability'': $\beta_p$ with $\alpha=0$ is the most regular
(and corresponds to the fully integrable case), whereas $\beta_s$ with
$\alpha \ne 0$ is the most ``nonintegrable''.

\begin{figure}[tpb]
\includegraphics[width=0.45\textwidth,clip=]{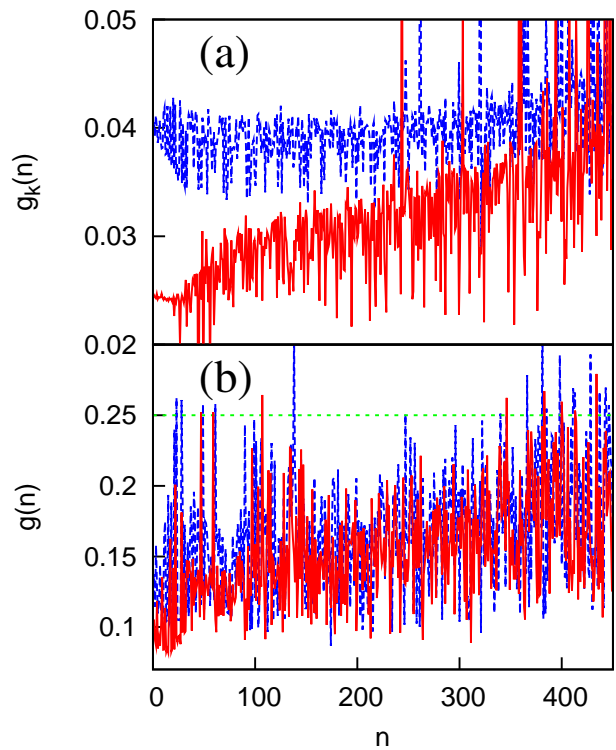} 
\caption{ \label{fig:localiz} (Color online) Localization $g_k$ (in
  momentum space) (a) and $g$ (in coordinate space) (b) of the
  eigenfunctions of $\beta_s$ for $\alpha=\pi/15$ (red solid line) and
  for $\alpha=0$ (blue dashed line) arranged according increasing of
  the absolute values of the imaginary parts of corresponding
  eigenvalues (i.~e. losses). In (b) $g$ is also shown for $\beta_p$
  and $\alpha=0$ (dotted green line). }
\end{figure}

As follows from the previous, the shape of eigenmodes can provide a
sensitive tool for the description of deviation of the system from the
full integrability.  This is supported by consideration of the
statistics of eigenfunction amplitudes $P(|\vect E|^2)$ shown in inset
to \reffig{fig:stat}. $P(|\vect E|^2)$ for $\beta_s$ for both $\alpha
\ne 0$ and $\alpha =0$ (they are very simular to each other and shown
by a single red curve in the inset) is sufficiently different to the
statistics of $\beta_p$ (blue curve).  Remarkably, $P(|\vect E|^2)$
for $\beta_s$ is very close to Porter-Thomas distribution $P(|\vect
E|^2) = e^{-|\vect E|^2/2}/\sqrt{2\pi |\vect E|^2}$ (which is
characteristic for chaotic billiards \cite{stoeckmann99}).

In this paper however, we are interested in analysis not only of
different ``levels of integrability'', but also of the localization
properties of eigenfunctions.  As it was shown in
\cite{backer99,chen03d,huang02}, sometimes the good indication of SLS
is a localization in the ``momentum'' space $\vect k_\bot$: $g_k =
\Delta H_0 /\langle H_0\rangle$, where $H_0 = k_\bot^2$, $\Delta H_0 =
\sqrt{\langle H_0^2 \rangle - \langle H_0 \rangle^2 }$.  $g_k$ becomes
exactly zero for the integrable case and is shown in
\reffig{fig:localiz}(a) for $\beta_s$. One can see that $g_k$ is
minimal at threshold ($n\sim 1$) \emph{in average} for $\alpha \ne 0$,
and grows with increasing losses ($n \rightarrow \infty$).  For ``less
nonintegrable'' case of $\alpha = 0$ the localization level is nearly
constant for low $n$, but starts to grow for $n$ above $\sim 200$.  In
general, $g_k$ can be used to distinguish between SLS and
randomly-distributed eigenfunctions in chaotic billiards
\cite{backer99}, but is not very well suitable for quasi-integrable
billiards, because the modes unlocalized in coordinate space can be
well localized in momentum space \cite{chen09}.

In this letter, we consider also a measure of localization in
coordinate space. Namely, we introduce a functional $g[I(x,y)] = \int
I \,dxdy/S I_{max}$ on the spatial intensity distribution $I(x,y) =
|\vect E(x,y)|^2$ (here the integration is made over the whole
billiard area, and the result is normalized to the area $S$ and to the
maximal value of the intensity $I_{max} = \mathrm{max}_{\vect
  r_\bot}\left(I(\vect r_\bot)\right)$.  $g$ varies from one (for
$I=\mathrm{const}$) to zero (for a pattern localized near a single
point, i.~e. for the one close to $\delta$-function).  Remarkably, for
an integrable square billiard $g=1/4$ for every eigenfunction (see
\reffig{fig:localiz}(b), gree dashed curve). Therefore, deviation of
$g$ from this value may indicate a deviation of the system from the
complete integrability.  In the other limit, for fully chaotic case,
the typical value of $g$ is rather small. One can estimate it using
the fact that eigenfunctions in fully chaotic billiards can be
simulated as a sum of the waves with similar $k_\bot$ but random
phases \cite{stoeckmann99}. For such a sum, according to our
computations, the value of $g$ averaged over large number of
eigenfunctions depends only on $k_\bot$ and is in the range of
$0.06-0.07$ for $k_\bot$ used in the present article.

The values of $g$ for $\beta_s$ and $\beta_p$ are shown in
\reffig{fig:localiz}.  For the case of $\beta_s$ the deviation of $g$
from the ``integrable'' value $0.25$ is quite strong.  Intriguingly,
in analogy to $g_k$ the average value of $g$ grows with $n$.  Thus,
SLS appearing at threshold can be described as the patterns having
lowest $g$ and $g_k$ simultaneously, supporting their consideration as
coherent states \cite{chen02a,chen03a,chen03d}.  In general, the whole
pair $\{g,g_k\}$ can be considered as a useful measure, allowing to
distinguish between SLS, delocalized as well as ``chaotic''
structures.

It should be noted that the deviation from the full integrability in
VCSEL disappears in a circular geometry \cite{gensty05}, because the
terms $s$ and $\Upsilon$ in $\beta_\infty$ are isotropic in that case.
In this sense, the situation is similar to the case of spin-orbit
coupling of electrons in quantum dots \cite{berggren01}. 

From the experimental point of view, it is rather problematic to
distinguish clearly between the effects, appearing due to small
deviations of the boundary conditions from a perfect square, and due
to deviation of operator of the system $\hat H$ from the
Laplacian. The result of the present article shows that the role of
the former is often overestimated.

Whereas localization in the momentum space has a concrete physical
reason in the case of VCSEL (certain $\vect k_\bot$ have the highest
gain because their frequencies are closer to the gain maximum),
localization in the coordinate space is not obvious from the physical
point of view. Even localization in the momentum space depends on the
level of integrability of the system (regulated by $\alpha$), although
the change of $\alpha$ does not affect the losses directly.

An intriguing similarity to the results of \cite{wiersig06} should be
pointed out, despite the mechanism of emission in VCSEL is completely
different from the one in dielectric microcavities. In
\cite{wiersig06}, long lived states created by connection of the
neighboring modes via continuum of external ones (i.~e. mediated by
dissipation) in the vicinity of ALC were shown to be localized in
coordinate space.  In contrast to \cite{wiersig06}, appearance of SLS
in our system is related only to PTC and not to the presence of
dissipation. Namely, according to our numerical simulations, if we
remove all the losses in the system (i.~e. assume $\kappa =0$ in
$\beta_\infty$), SLS \textit{do not} disappear. Moreover, the shape of
many SLS is not altered significantly. Thus, the presence of
dissipation does not create SLS but orders them instead.  In addition,
SLS are also not directly related to ALC in our case, because the
frequencies of the modes grow in average monotonically with their
losses, which is not the case in the vicinity of ALC.

In general, the results of this Letter together with \cite{wiersig06,
  ermann09} allows us to suspect a general mechanism, which is
independent from every particular physical realization of open
billiard, leading to a relation between localization and losses, which
is still to be clarified.

 The author is grateful for useful discussions with T. Ackemann.


%
    
\end{document}